\documentclass[10pt,conference]{IEEEtran}
\IEEEoverridecommandlockouts
\usepackage{balance}
\usepackage{amssymb}
\usepackage{stmaryrd}
\usepackage{cite}
\usepackage{color}
\usepackage{multirow}
\usepackage{graphicx,times}
\usepackage{epstopdf}
\usepackage{indentfirst}
\usepackage{CJK}
\usepackage{amsmath}
\usepackage{amsfonts}
\usepackage{txfonts}
\usepackage{mathrsfs}
\usepackage{subfigure}
\usepackage{graphicx}
\usepackage{url}
\usepackage{fancyhdr}   

\newtheorem{proposition}{Proposition}

\newtheorem{corollary}{Corollary}

\newtheorem{remark}{Remark}

\begin{document}
\title{Towards Motion Invariant Authentication for On-Body IoT Devices}
\author{\IEEEauthorblockN{Yong Huang{$^{^\dagger}$}, Mengnian Xu{$^{^\dagger}$}, Wei Wang{${^\ast}{^{^\dagger}}$}, Hao Wang{$^{^\ddagger}$}, Tao Jiang{$^{^\dagger}$}, Qian Zhang{${^\S}$}}\IEEEauthorblockA{{$^{^\dagger}$}School of Electronic Information and Communications, Huazhong University of Science and Technology\\ {$^{^\ddagger}$}  Computer Science and Artificial Intelligence Lab, Massachusetts Institute of Technology \\{${^\S}$} Department of Computer Science and Engineering, Hong Kong University of Science and Technology\\Email: \{yonghuang, mengnian, weiwangw, taojiang\}@hust.edu.cn, hwang87@mit.edu, qianzh@cse.ust.hk}
\thanks{${^\ast}$The corresponding author is Wei Wang (weiwangw@hust.edu.cn).}
}	

\maketitle
\thispagestyle{fancy}          
\fancyhead{}                     
\chead{IEEE ICC 2019 - Communication and Information Systems Security Symposium}
\cfoot{} 
\renewcommand{\headrulewidth}{0pt}     
\renewcommand{\footrulewidth}{0pt}

\begin{abstract}
As the rapid proliferation of on-body Internet of Things (IoT) devices, their security vulnerabilities have raised serious privacy and safety issues. Traditional efforts to secure these devices against impersonation attacks mainly rely on either dedicated sensors or specified user motions, impeding their wide-scale adoption. This paper transcends these limitations with a general security solution by leveraging ubiquitous wireless chips available in IoT devices. Particularly, representative time and frequency features are first extracted from received signal strengths (RSSs) to characterize radio propagation profiles. Then, an adversarial multi-player network is developed to recognize underlying radio propagation patterns and facilitate on-body device authentication. We prove that at equilibrium, our adversarial model can extract all information about propagation patterns and eliminate any irrelevant information caused by motion variances. We build a prototype of our system using universal software radio peripheral (USRP) devices and conduct extensive experiments with both static and dynamic body motions in typical indoor and outdoor environments. The experimental results show that our system achieves an average authentication accuracy of 90.4\%, with a high area under the receiver operating characteristic curve (AUROC) of 0.958 and better generalization performance in comparison with the conventional non-adversarial-based approach. 
\end{abstract}

\maketitle

\section{Introduction}

With continuing advances in sensors and low-power communication technologies, human-centric Internet of Things (IoT) has gained increasing momentum in both industrial and academic communities by unobtrusively providing smart user-centered services \cite{uckelmann2011architectural,mainetti2015iot}. The hardware miniaturization of IoT devices resembles the two sides of a coin: it empowers IoT devices to communicate via ultra low-power radios while making communication links vulnerable to malicious invasions. Since on-body IoT devices are generally attached to users' bodies to continuously record fine-grained vital signs, security breaches of these devices pose a serious threat to users' everyday privacy and safety \cite{gollakota2011they}.

Growing attempts and extensive endeavors have been devoted to thwarting malicious masqueraders for hardware-constrained wearable devices. It has been shown that radio channel characteristics in body area networks (BANs) can be exploited to perform device authentication \cite{shi2013bana}. Recent efforts have also leveraged dedicated sensors \cite{revadigar2017accelerometer,xu2017gait,cornelius2014wearable,vu2012distinguishing}, such as accelerometers and gyroscopes, to verify wearable devices. However, hardly any of them have achieved widespread acceptance. They limit themselves to either special motion scenarios \cite{shi2013bana}, or fitness related wearables \cite{revadigar2017accelerometer,xu2017gait,cornelius2014wearable,vu2012distinguishing}. To embrace the coming wave of human-centric IoT, it is critical for a device authentication solution to support various on-body IoT devices under diverse user motions.

The salient physical layer (PHY) signatures underlying different BANs present us with an exciting opportunity. In BAN channels, off-body signals are mainly comprised of line-of-sight (LOS) and multi-path components, while on-body signals are governed by \textit{creeping waves} \cite{di2011body,ryckaert2004channel}. The distinct radio propagation patterns potentially enable a general security solution relying on prevalent wireless chips. However, radio signals in BAN channels are severely affected by IoT users' body motions. As a consequence, on- and off-body signals can exhibit significantly different patterns under a specific user motion, and their patterns tend to vary dramatically across multiple motion states. Furthermore, users' frequent motion changes in daily life make it a highly challenging task to manually select features to represent propagation patterns from real-world radio traces. 

To address this challenge, we propose a motion invariant authentication framework for on-body IoT devices. The proposed system performs device authentication by exploiting BAN radio signatures in two steps. In the first step, our system abstracts representative time and frequency features from noisy received signal strength (RSS) segments to characterize fine-grained radio propagation characteristics. In the second step, to learn robust feature representations from abstracted radio features, an adversarial multi-player network is customized to effectively remove motion specific features and thereafter accurately recognize the identities of IoT devices. To achieve this goal, during training, an adversarial training criterion is implemented, which leads to the emergence of transferable features that generalize well in unseen motion states. We implement a working prototype of our system on universal software radio peripheral (USRP) devices and conduct experiments with various body motions in different real-world environments. Experimental results show that our system achieves an authentication accuracy of 90.4\% on average.

The main contributions of this work are summarized as follows.
\begin{itemize}
	\item We propose a general authentication system that supports various on-body IoT devices under diverse body motions. The crux of the proposed system is to construct reliable radio propagation profiles from RSS segments and to develop an adversarial network to essentially identify IoT devices based on underlying propagation patterns.
    \item We theoretically analyze our adversarial multi-player network and demonstrate that at equilibrium, the learned feature representation contains all information about BAN radio propagation patterns, and becomes invariant to user body motions.
    \item We build a prototype of our system on USRP platform and conduct extensive experiments with various frequently appearing body motions in a variety of indoor and outdoor environments. The experimental results demonstrate the effectiveness and generalizability of our system.
\end{itemize}

\section{Exploiting Distinct Radio Propagation Patterns between On- and Off-Body Channels}

\begin{figure}
	\centering
	\includegraphics[width=0.85\linewidth]{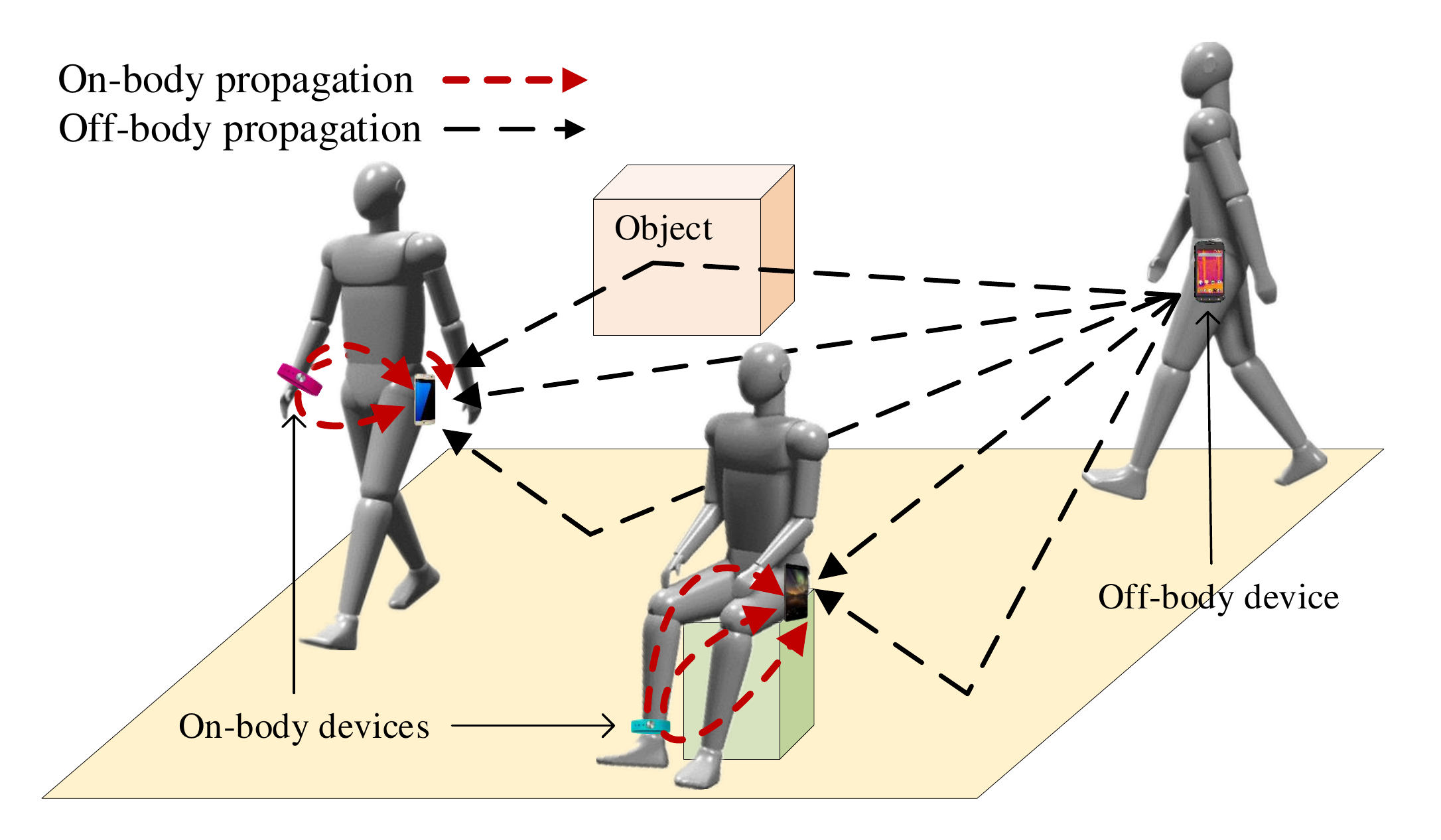}
	\caption{On- and off-body radio propagations. On-body signals are dominated by creeping waves, and off-body signals are mainly comprised of LOS and multi-path components.}
	\label{fig:propagation-patterns}
\end{figure}

Since the human body is basically a low-loss dielectric at microwaves frequencies, including Wi-Fi and Bluetooth frequency bands, radio propagation between a transmitter (Tx) and a receiver (Rx) carried by a user is significantly influenced by the user's body. As shown in Fig.~\ref{fig:propagation-patterns}, off-body links are dominated by LOS and multi-path propagations. On the other hand, on-body links are governed by \textit{creeping waves}, which are diffracted by human tissues and spread out along the human body \cite{ryckaert2004channel}. Previous measurements \cite{di2011body} indicate that creeping waves are rarely disturbed by multi-path fading (small-scale fading) or large-scale fading caused by Tx-Rx distance changes or shadowing, but are largely influenced by body motions. Thus, we see that distinct propagation patterns exist between on- and off-body signals.

\begin{figure}
	\centering
	\subfigure[Signals in the standing state.]{
		\includegraphics[width=0.40\linewidth]{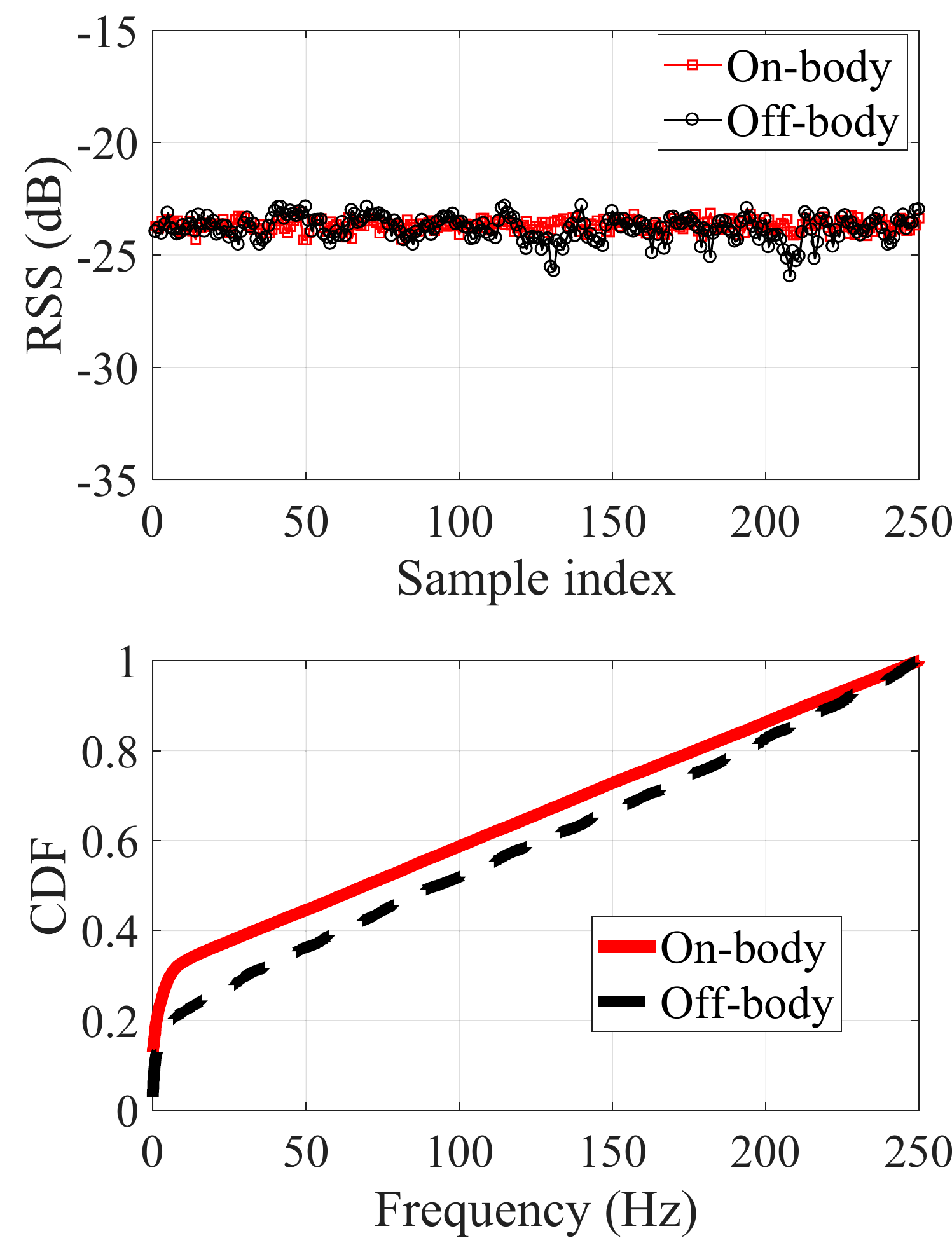}}
	\label{1a}\hfill
	\subfigure[Signals in the walking state.]{
		\includegraphics[width=0.40\linewidth]{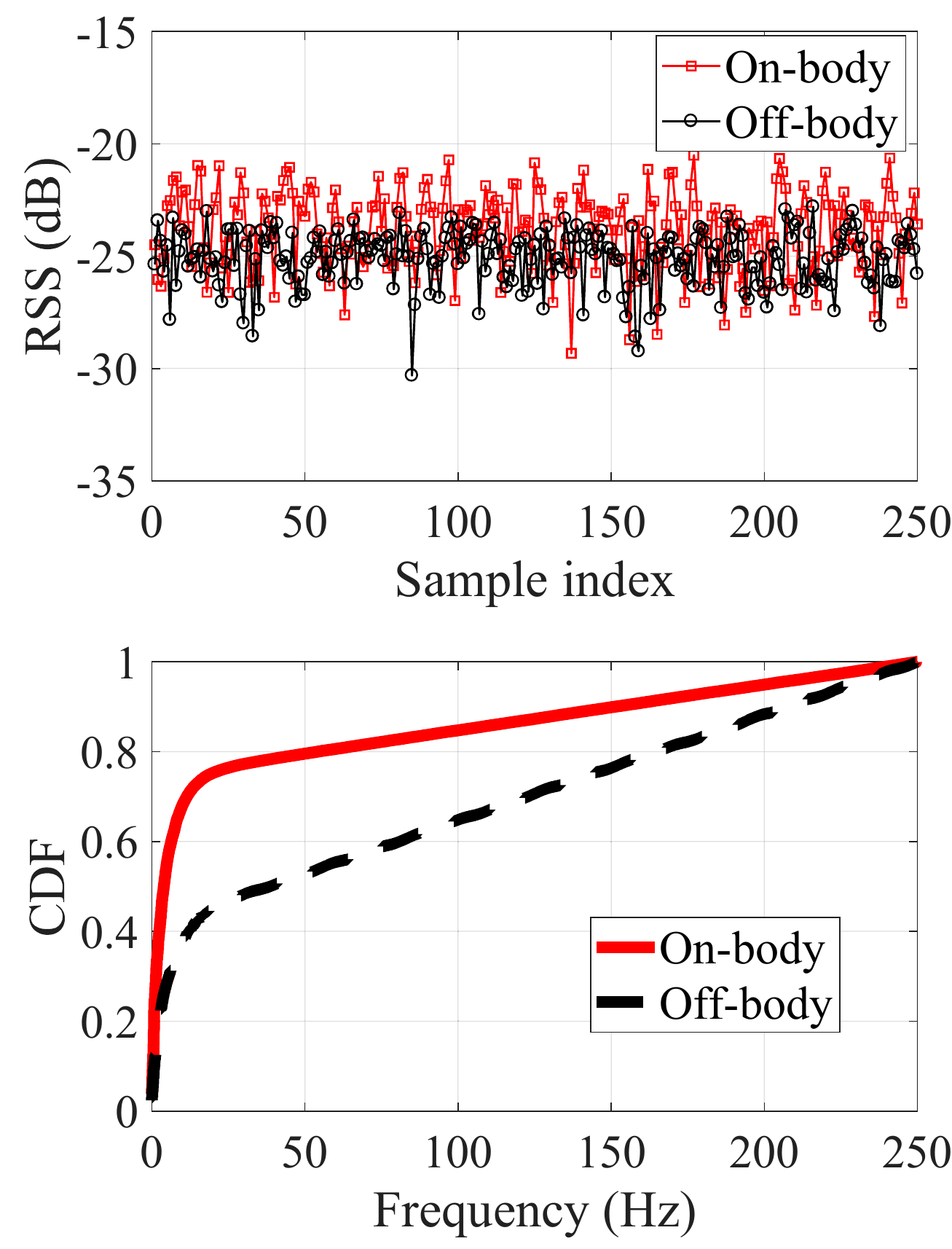}}
	\label{1b}\\
	\caption{RSS and CDF of on- and off-body radio signals in standing and walking states.}
	\label{comparison between static and dynamic motion} 
\end{figure}

Fig.~\ref{comparison between static and dynamic motion} depicts the RSS and the cumulative distribution function (CDF) of different BAN signals that were collected in standing and walking states, respectively. In the standing state, we observe that on-body signals are more stable in the time domain and off-body signals contain more components in the high frequency band. In the walking state, on-body signals have a larger RSS variance and fall into a low frequency range with a very high probability. The experimental observations verify that differentiable propagation patterns exist in on- and off-body channels in each motion state. This supports our premise that we can rely upon PHY signatures to authenticate various on-body IoT devices. 

\section{Adversarial Network Based Device Authentication}

\subsection{Design Rationale}
It is, however, non-trivial to reliably capture propagation patterns from real-world radio traces. As shown in Fig.~\ref{comparison between static and dynamic motion}, although on- and off-body signals show distinguishable propagation patterns in each case, their patterns are remarkably different between the two cases. Consequently, an authentication model that is trained under a specific user motion will typically not generalize well in different motion scenarios. 

To deal with such dilemma, we resort to adversarial networks, which have recently surfaced as a popular tool to discover transferable features in the deep learning field and have proven their advantages in many real-world applications \cite{ganin2016domain,zhao2017learning,shinohara2016adversarial}. Being a branch of deep learning approaches, adversarial networks facilitate automatic extraction of complex and latent feature representations by adopting a hierarchical structure \cite{goodfellow2016deep}. Different from traditional approaches, they have the ability to find and exclude irrelevant features in the learned representations with an adversarial training criterion. 

Therefore, we can reap the benefits of adversarial networks to recognize underlying on- and off-body propagation patterns. In our application, a customized adversarial network can be leveraged to autonomously extract feature representations about BAN radio propagation patterns and selectively eliminate motion specific features from the representations. To this end, we propose an adversarial network based security system to seamlessly authenticate various on-body IoT devices.

\subsection{Design Overview}
\begin{figure}
	\centering
	\includegraphics[width=0.9\linewidth]{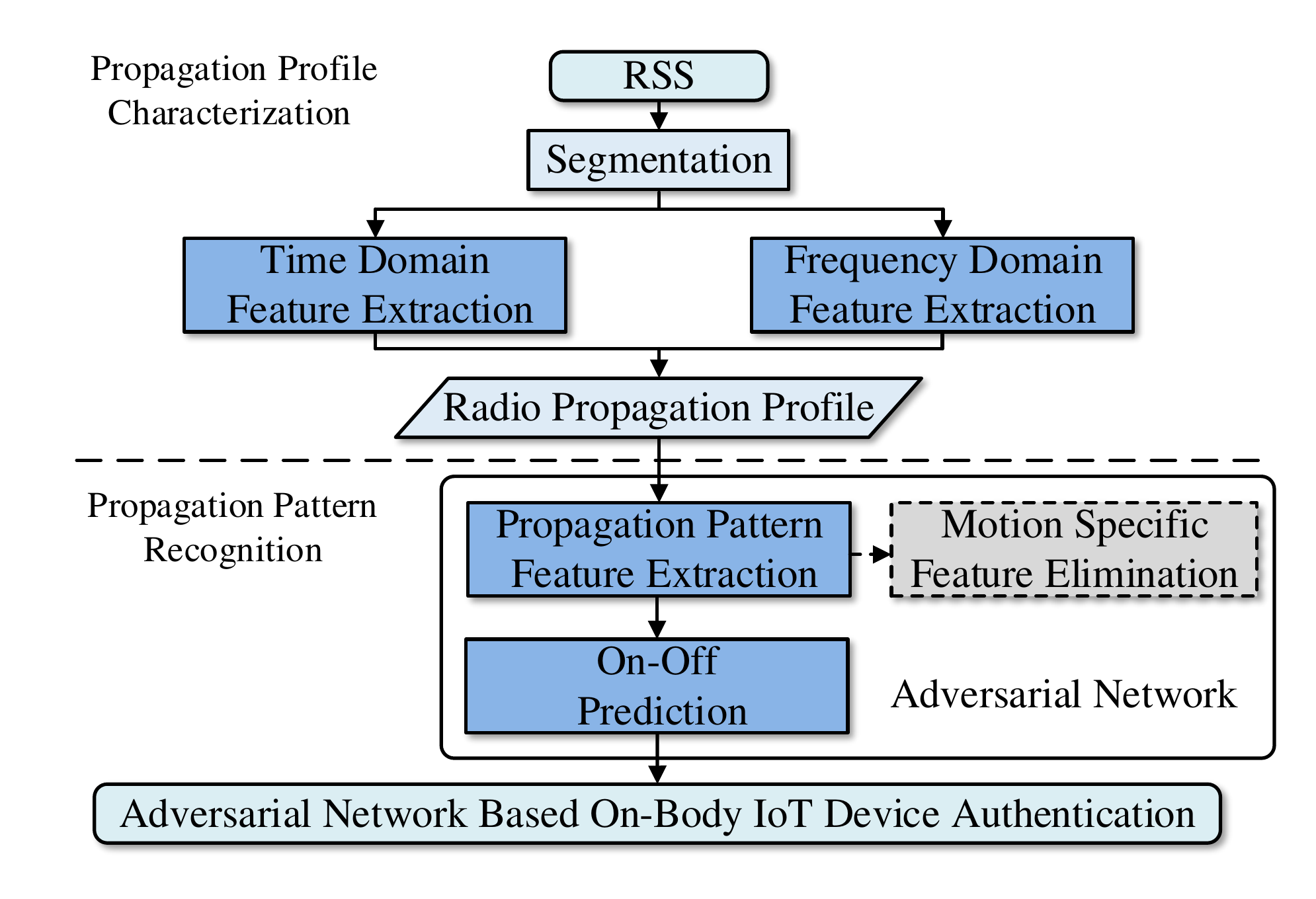}
	\caption{System flow. The dashed gray block exists only in the training phase.}
	\label{fig:system-flow}
\end{figure}

Our system takes advantage of an adversarial network to extract distinct radio propagation patterns for on-body device authentication. Fig.~\ref{fig:system-flow} illustrates the framework of our authentication system. It takes as input RSS time series and outputs the corresponding device authentication results. It is worth noting that to verify RSSs from various low-end embedded IoT devices, our authentication system locates at users' smartphones, which have sufficient capability to perform low-latency and accurate learning based inferences \cite{mobilenet}. 

The core of our authentication system includes two components -- \textit{Propagation Profile Characterization} and \textit{Propagation Pattern Recognition}.
\begin{enumerate}
    \item  \textbf{Propagation Profile Characterization.} This component first divides RSS time series into multiple basic segments. Then, radio features are extracted from both the time and frequency domains of RSS segments for fine-grained characterization of potential propagation patterns. Finally, the extracted features are integrated into radio propagation profiles for future pattern recognition by the adversarial network. 
	\item  \textbf{Propagation Pattern Recognition.} Upon receiving a propagation profile, the adversarial network first utilizes a functional block to abstract a feature representation in terms of on- and off-body propagations. Subsequently, the model infers the identity of a connected IoT device through an on-off prediction block. Moreover, an adversary block is added to eliminate motion specific features in the feature representation in the training phase. All blocks are learned through an adversarial training process to promote the emergence of features that are resilient to motion changes. 
\end{enumerate}

\subsection{Propagation Profile Characterization}

\textbf{Signal Segmentation.} Our system first partitions RSS measurements into multiple segments. As an RSS segment is the basic unit for device authentication, the segment interval needs to be carefully determined. If the interval is too long, on- and off-body signals will be probably both included in a same segment. If it is too short, the system will be unable to recognize any segment. We empirically find that a time interval of 5s is capable of correctly differentiating over $ 90\% $ of on- and off-body IoT devices.

\textbf{Time Domain Feature Extraction.} Since on- and off-body signals have different levels of impact from body motions, large- and small-scale fading, we first decompose each RSS segment into multi-scale variations by using filters. As creeping waves are sensitive to body motions and their frequencies fall into relatively low frequency bands with a high probability \cite{xiong2006hand}, a band-pass filter is leveraged to extract motion-induced variations. Based on our experimental observations, most fluctuations caused by body motions fall between 0.5 Hz and 15 Hz. Variations in the residual low and high frequency bands are also extracted by a low-pass filter and a high-pass filter, respectively, as large- and small-scale variations.

With multi-scale variations, we select six time domain features, including \textit{maximum}, \textit{minimum}, \textit{median}, \textit{variance}, \textit{kurtosis} and \textit{skewness}, to characterize propagation signatures from each kind of variations. The maximum, minimum, median and variance are chosen to describe the impact from the human body, because dramatic body vibration typically contributes to rapid changes in the maximum, minimum and median and also results in a large variance. Kurtosis and skewness show the symmetry and asymmetry of radio signals, respectively, and can potentially capture propagation patterns due to the fact that both symmetric and asymmetric components are richly shared in radio waves. For finer-grained feature extraction, we divide each kind of variations into ten chunks and extract six features from each chunk. Therefore, a total of 180 feature points are extracted to describe radio propagation signatures from the time domain of an RSS segment.

\textbf{Frequency Domain Feature Extraction.} To abstract frequency domain features, we start by performing Short-Time Fourier Transform (STFT) on each RSS segment to obtain its two-dimensional spectrogram. Specifically, with a signal sampling rate of 500 Hz, we conduct a 1000-point Fast Fourier Transform (FFT) within a 2s sliding window, shifting 1s each time to make full use of sampling data. To summarize information in the frequency domain, the frequency band of each spectrogram, i.e., [0,250] Hz, is partitioned into 40 intervals, each of which is associated with a frequency component of the segment. To effectively indicate propagation signatures, we equally segment the low frequency band, i.e., [0,15] Hz, into 30 intervals and the residual high frequency band into 10 intervals, and we sum up the magnitudes in each interval in every FFT result. In this way, we transform a two-dimensional spectrogram into a 4$ \times $40 matrix $ \mathbf{M} $. Then we take two frequency domain features from $ \mathbf{M} $: the \textit{component magnitude} (or each element in $ \mathbf{M} $) and the \textit{proportion of each component} (PC), such that $ PC(j)=\frac{\sum_{i=1}^{4} \mathbf{M}(i,j)}{\sum_{j=1}^{40} \sum_{i=1}^{4} \mathbf{M}(i,j)} $, where $ j=1, \cdots, 40$. Finally, a total of 200 feature points are extracted from the frequency domain of an RSS segment. 

\subsection{Propagation Pattern Recognition}
We formulate the propagation pattern recognition as a binary classification task, where $ \mathcal{X} \subseteq \mathbb{R}^{n} $ is the sample space and $ \mathcal{Y} = \left\lbrace 0,1 \right\rbrace  $ is the target label set. In our context, each $ \mathbf{x} \in \mathcal{X} $ is a radio propagation profile sample, and $ y \in \mathcal{Y} $ indicates the corresponding on- or off-body IoT device. Moreover, for each $ \mathbf{x} $, $ z \in \left\lbrace 1, 2, \cdots, n_{z} \right\rbrace  $ denotes an auxiliary label that refers to the body motion that $ \mathbf{x} $ is sampled from. 
\begin{figure}
	\centering
	\includegraphics[width=0.9\linewidth]{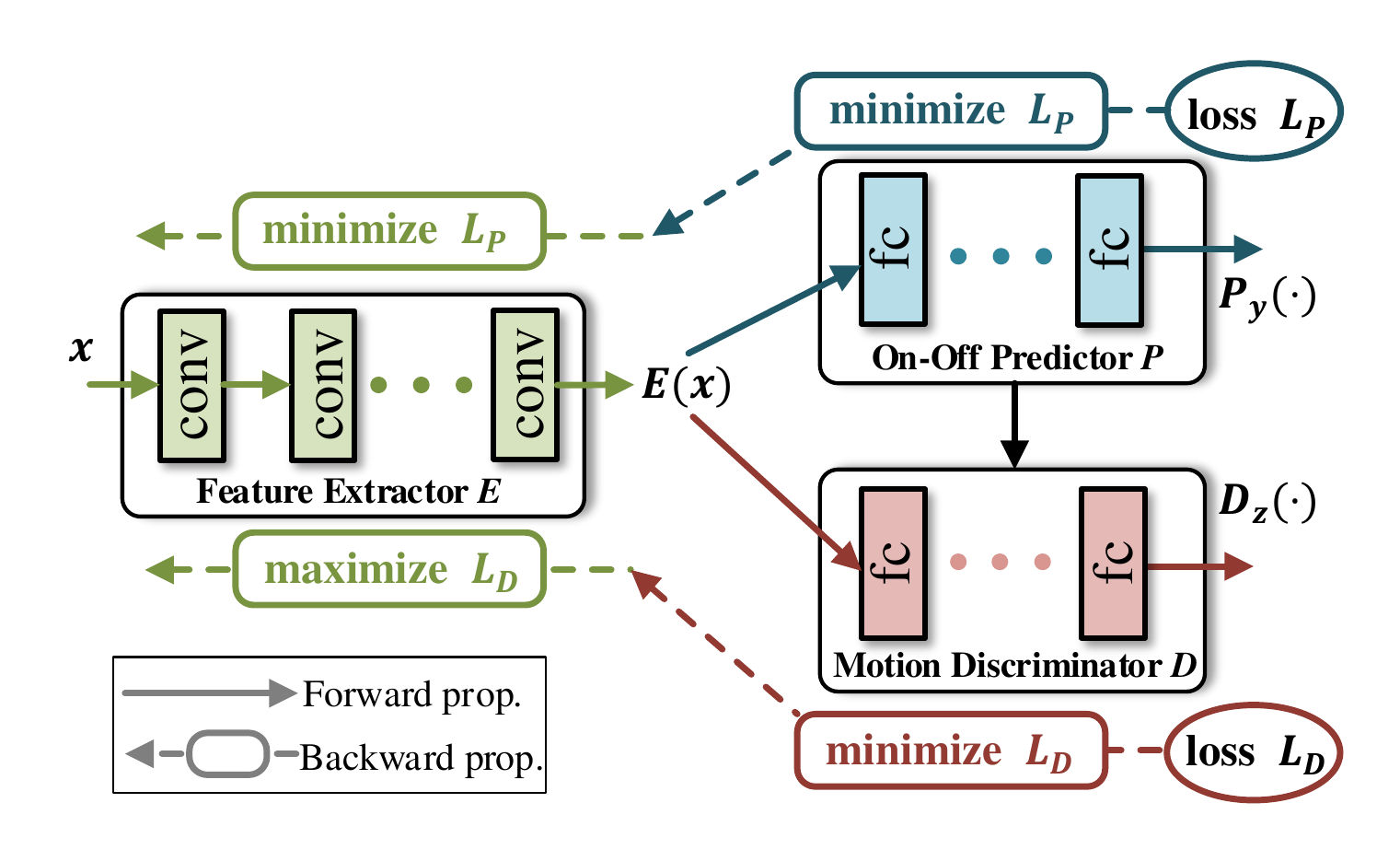}
	\caption{Our adversarial multi-player network and corresponding adversarial training criterion. The model encompasses three blocks -- a Feature Extractor $ E $, an On-Off Predictor $ P $ and a Motion Discriminator $ D $.}
	\label{fig:adversarial-model}
\end{figure}

We develop an adversarial multi-player network for propagation pattern recognition. As depicted in Fig.~\ref{fig:adversarial-model}, our model encompasses three blocks -- a \textit{Feature Extractor $ E $}, an \textit{On-Off Predictor $ P $} and a \textit{Motion Discriminator $ D $}. Since simply wiping out all dependencies between feature representations and domains (e.g., motions in our application) could degrade the accuracy of target label prediction, our model adopts a conditional adversarial architecture\cite{zhao2017learning} for better generalization performance.

\textbf{Feature Extractor $ E $.} An extractor $ E $ is the front block of the adversarial model. It takes as input a propagation profile $ \mathbf{x} $ and returns a latent feature representation as $ E(\mathbf{x}) $. 

\textbf{On-Off Predictor $ P $.} A predictor $ P $ acts as the end block of the model. It takes as input a learned feature representation $ E(\mathbf{x}) $ and outputs a two-dimensional probability vector $ P_{y}(\cdot|(E(\mathbf{x}))) $ in terms of on- and off-body devices.

\textbf{Motion Discriminator $ D $.} A discriminator $ D $ serves as an adversary in our model in the training phase. It takes a feature representation $ E(\mathbf{x}) $ and the associated on-off probability vector $ P_y (\cdot|E(\mathbf{x})) $ together as input, and discriminates which motion state $ \mathbf{x} $ is sampled from as $ D_z \left(\cdot \big| E(\mathbf{x}), P_y (\cdot|E(\mathbf{x})) \right) $. 

\textbf{Adversarial Training.} 
Before performing authentication, our adversarial model needs to be trained on training data, which follows the distribution $ q_{train} (\mathbf{x}) $. We define the loss of $ P $ as the cross-entropy between $ P_{y} \left( \cdot|E(\mathbf{x})\right)  $ and the true posterior target label distribution $ q_y (\cdot|\mathbf{x}) $ over $ q_{train} (\mathbf{x}) $, which is given as
\begin{align}
\mathcal{L}_P (P,E) \triangleq \mathbb{E}_{\mathbf{x} \sim q_{train} (\mathbf{x})} \mathbb{E}_{y \sim q (y|\mathbf{x})} \left[ -\log P \left( y | E(\mathbf{x})\right) \right].
\end{align}
Similarly, the loss of $ D $ is defined as the cross-entropy between $ D_z \left(\cdot \big| E(\mathbf{x}), P_y (\cdot|E(\mathbf{x})) \right) $ and the true conditional distribution $ q_z (\cdot|\mathbf{x}) $ over $ q_{train} (\mathbf{x}) $, which is expressed as
\begin{align}\label{loss of d}
\mathcal{L}_{D} \left( D,E;P\right) \triangleq \mathbb{E}_{\mathbf{x} \sim q_{train} (\mathbf{x})} \mathbb{E}_{z \sim q (z|\mathbf{x})} \left[ -\log D \left( z \big| E(\mathbf{x}),P_y (\cdot|E(\mathbf{x})) \right) \right].
\end{align}
Note that to effectively learn parameters of our multi-player model, the flow from $ P $ to $ D $ is a one-way link (i.e., the black arrow line in Fig.~\ref{fig:adversarial-model}), along which gradients don't propagate back. Thus, the parameters of $ P $ are not updated through the optimization of the loss $ \mathcal{L}_{D}$. 

To robustly authenticate IoT devices under diverse body motions, it is critical for our model to implement an adversarial training criterion. The basic idea is that to generalize well in unseen scenarios, a predictive model must discriminate well between on- and off-body devices, but it cannot distinguish body motions associated with input samples. To achieve this goal, we use minimax games between $ E $, $ P $ and $ D $ in the training phase. Particularly, $ E $ plays a cooperative game with $ P $ to minimize the loss $ \mathcal{L}_P $. At the same time, $ E $ and $ D $ together play a minimax game, where $ D $ aims to minimize the loss $ \mathcal{L}_{D} $ and $ E $ tries to maximize it. 

We integrate the above objectives into one value function: 
\begin{align}\label{value function}
\mathcal{V}\left( E,P,D\right)  \triangleq & \mathcal{L}_P (P,E) - \lambda \cdot \mathcal{L}_{D} (D,E; P),
\end{align}
where $ \lambda > 0 $ is hyperparameter. With the value function \eqref{value function}, the adversarial training criterion can be implemented by optimizing the following minimax problem:
\begin{align}\label{minimax training criterion}
\mathop{\min} \limits_{E,P} \mathop{\max} \limits_{D} \mathcal{V}\left( E,P,D\right).
\end{align}

\subsection{Theoretical Analysis of Adversarial Model}
We prove that the output of our adversarial model becomes invariant to motion changes through the adversarial training. Specifically, we first present the optimal predictor and optimal discriminator in Proposition 1 and Proposition 2, respectively, without proving them, and refer the reader to \cite{zhao2017learning} (Proposition 2) for details. Then, we illustrate the virtual training criterion, optimal extractor and optimal output, respectively, in Corollary 1, Proposition 3 and Corollary 2. Differing from the theoretical efforts in the prior work \cite{zhao2017learning}, our analysis focuses on a practical adversarial model.
\begin{proposition} 
	(Optimal predictor) For a fixed extractor $ E $, the output of the optimal predictor $ P^{*} $ over $ q_{train} (\mathbf{x}) $ achieves
	\begin{align} \label{optimal predictor}
	P^{*}\left( y|E(\mathbf{x})\right) =q(y|E(\mathbf{x})),
	\end{align}	
	and the loss of $ P^{*} $ is
	\begin{align} \label{loss of optimal predictor}
	\mathcal{L}_{P^{*}} (E) \triangleq \mathop{\min} \limits_{P} \mathcal{L}_P (P,E) = H(y|E(\mathbf{x})),
	\end{align}
	where $ H(\cdot| \cdot) $ denotes the conditional entropy function.
\end{proposition}

Note that given $ E $, the equality \eqref{optimal predictor} indicates the maximal predictive capability that a predictor $ P $ can learn from $ q_{train} (\mathbf{x}) $.
\begin{proposition} 
	(Optimal discriminator) Given any extractor $ E $ and any predictor $ P $, the optimal discriminator $ D^{*} $ over $ q_{train} (\mathbf{x}) $ obtains
	\begin{align} \label{optimal D1}
	D^{*} \left( z \big| E(\mathbf{x}), P_y (\cdot|E(\mathbf{x})) \right) = q\left( z \big| E(\mathbf{x}),P_y (\cdot|E(\mathbf{x})) \right),
	\end{align}
	and its loss is
	\begin{align} \label{loss of optimal D1}
	 \mathcal{L}_{D^{*}}(E;P)  \triangleq \mathop{\min} \limits_{D} \mathcal{L}_{D}(D,E;P)
	 = H\left( z \big|E(\mathbf{x}), P_y (\cdot|E(\mathbf{x})) \right).
	\end{align}
\end{proposition}

With the optimal predictor and optimal discriminator, we proceed to simplify the minimax training criterion \eqref{minimax training criterion}.

\begin{corollary} (Virtual training criterion)
	If $ P $ and $ D $ have enough capacity and are trained to be optimal over $ q_{train} (\mathbf{x}) $, the minimax optimization \eqref{minimax training criterion} is equivalent to the minimization of a virtual value function $ \mathcal{V}(E) $, which is expressed as  
	\begin{align} \label{actual minimax training criterion}
    \mathcal{V}(E) \triangleq H \left( y |E(\mathbf{x}) \right) - \lambda \cdot H\left( z \big| E(\mathbf{x}),q_y(\cdot|E(\mathbf{x}) )  \right). 
	\end{align}
\end{corollary}

\begin{IEEEproof}
	Considering the optimal predictor $ P^{*} $ in Proposition 1, we can rewrite the loss of the optimal discriminator $ D^{*} $ in Proposition 2, by substituting \eqref{optimal predictor} into \eqref{loss of optimal D1}, as
	\begin{align} \label{new loss of optimal D}
	\mathcal{L}_{D^{*}}(E) = H\left( z \big|E(\mathbf{x}),q_y (\cdot|E(\mathbf{x})) \right).
	\end{align}
	According to the losses of the optimal predictor \eqref{loss of optimal predictor} and optimal discriminator \eqref{new loss of optimal D}, the initial value function \eqref{value function} can be simplified as the virtual version \eqref{actual minimax training criterion}. Thus, optimizing the minimax optimization \eqref{minimax training criterion} equals to minimizing $ \mathcal{V}(E) $. 
\end{IEEEproof}

Then, we obtain the optimal extractor by minimizing $ \mathcal{V}(E) $. 

\begin{proposition} (Optimal extractor) \label{optimal extractor} 
	 If $ E $, $ P $ and $ D $ have enough capability and are trained to be optimal over $ q_{train} (\mathbf{x}) $, any optimal extractor $ E^{*} $ satisfies 
	\begin{align}\label{property_one}
	H \left( y| E^{*} (\mathbf{x}) \right) = H \left( y|\mathbf{x} \right), 
	\end{align}
	and
	\begin{align}\label{property_two}
	H \left( z \big| E^{*} (\mathbf{x}), q_y (\cdot|E^{*} (\mathbf{x})) \right) = H \left( z \big|  q_y (\cdot|E^{*} (\mathbf{x})) \right). 
	\end{align}
\end{proposition}
\begin{IEEEproof}
	When $ E $ is fixed, $ \mathcal{L}_{P^{*}} (E) = H \left( y|E(\mathbf{x}) \right) \ge H(y|\mathbf{x}) $ and $ \mathcal{L}_{D^{*}} (E) = H \left( z \big| E(\mathbf{x}), q_y (\cdot|E(\mathbf{x})) \right) \le H \left( z \big| q_y (\cdot|E(\mathbf{x})) \right) $.
	Therefore, we obtain a lower bound of $ \mathcal{V}(E) $, that is
	\begin{align}
	\mathcal{V}(E) \ge  H(y|\mathbf{x})- \lambda \cdot H \left( z \big| q_y (\cdot|E(\mathbf{x})) \right).
	\end{align}
	Since the bound is achieved if and only if both the conditions \eqref{property_one} and \eqref{property_two} hold, proving that any optimal extractor $ E^{*} $ satisfies \eqref{property_one} and \eqref{property_two} is identical to proving the equality $ \mathcal{V}(E^{*}) = H(y|\mathbf{x})- \lambda \cdot H \left( z \big| q_y (\cdot|E^{*}(\mathbf{x})) \right) $. 
	
	We note that the lower bound is achievable by considering a special case, where $ E^{*} (\mathbf{x}) = q_y (\cdot|\mathbf{x}) $, an extractor with the best representative ability. In this case, we can check that 
	\begin{align}
	\notag \mathcal{V}(E^{*}) & = H(y|E^{*}(\mathbf{x})) - \lambda \cdot H\left( z \big| E^{*} (\mathbf{x}),q_y (\cdot|E^{*} (\mathbf{x}))\right)\\
	& = H(y|\mathbf{x}) - \lambda \cdot H(z|q_y (\cdot|E^{*} (\mathbf{x}))).
	\end{align}
\end{IEEEproof}
\begin{remark}
	Proposition 3 indicates that when all blocks are trained to be optimal and our adversarial model reaches equilibrium, the extractor $ E $ is able to extract all information about $ y $ from the training samples and eliminate any information about $ z $ except what is also related to $ y $.
\end{remark}

\begin{corollary} (Optimal output)
	If $ E $, $ P $ and $ D $ have enough capacity and are trained to be optimal over $ q_{train} (\mathbf{x}) $, the output of our adversarial model achieves
	\begin{align}
	P_y (\cdot|E(\mathbf{x})) = q_y (\cdot|\mathbf{x}),
	\end{align}
	and
	\begin{align}
	D_z \left( \cdot \big| E(\mathbf{x}), P_y (\cdot|E(\mathbf{x})) \right) = q_z (\cdot| q_y (\cdot|\mathbf{x})).
	\end{align}
\end{corollary}
\begin{IEEEproof}
	Based on Proposition 1, $ P\left( y|E(\mathbf{x})\right) =q(y|E(\mathbf{x})) $. According to Proposition 3, $ H \left( y| E (\mathbf{x}) \right) = H \left( y|\mathbf{x} \right) $, which implies that $ q(y|E(\mathbf{x})) = q(y|\mathbf{x}) $. Hence, $ P (y|E(\mathbf{x})) = q(y|\mathbf{x}) $.
	
	When $ P $ and $ D $ are optimal, the equality \eqref{new loss of optimal D} holds, that is $ D \left( z \big| E(\mathbf{x}), P_y (\cdot|E(\mathbf{x})) \right) = q\left( z \big|E(\mathbf{x}), q_y (\cdot|E(\mathbf{x}))\right)  $. According to Proposition 3, the equality \eqref{property_two} holds, which is equivalent to $ q\left( z \big|E(\mathbf{x}), q_y (\cdot|E(\mathbf{x}))\right) = q(z|q_y (\cdot|E(\mathbf{x}))) $. Then, by considering $ P(y|E(\mathbf{x})) = q(y|\mathbf{x}) $, we achieve the equality $ D \left( z \big| E(\mathbf{x}), P_y (\cdot |E(\mathbf{x})) \right) = q(z|q_y (\cdot|\mathbf{x})) $. 
\end{IEEEproof}

\section{Evaluation in Real Environments}

\subsection{Experimental Methodology}
\textbf{Implementation.} We build a proof-of-concept prototype of the proposed system with three GNURadio/USRP B210 devices, which work at 2.4 GHz with a sampling rate of 500 Hz. Furthermore, two USRP devices are placed on a volunteer, referred to as a legitimate user, and are considered to be two on-body devices. The left device is situated on another volunteer, referred to as a malicious attacker, and is regarded to be an off-body device.

\textbf{Data Collection.} We collect radio traces in both controlled and uncontrolled user motion scenarios. In the controlled scenario, the user is confined to five frequently appearing motions, which are comprised of two static motions, \textit{sitting} and \textit{standing}, and three dynamic ones, \textit{arm moving}, \textit{rotating} and \textit{walking}. In the uncontrolled scenario, the user is permitted to behave casually. In both scenarios, the attacker is allowed to move freely in the vicinity of the user to try to fool the legitimate devices. Moreover, to verify the robustness of our system under various environments, we collect wireless signals in five indoor and outdoor settings, i.e., a lab, a meeting room, a corridor, a rooftop and a park. We conduct the experiments over seven days and collect a total of ten hours of radio traces.

\textbf{Dataset.} Our dataset includes a total of 7200 samples that are extracted from collected radio traces. Therein, 6000 samples are from the controlled user motion scenario, and 1200 are from the uncontrolled scenario. When evaluating our model, we randomly take out 4800 samples from the controlled scenario for training and combine the leftover 1200 ones and all 1200 samples from the uncontrolled scenario for testing. Additionally, in both the training and testing sets, the numbers of on- and off-body samples are equal.

\textbf{Parameterization.} As shown in Fig.~\ref{fig:adversarial-model}, we parameterize our multi-player model as a deep neural network. Specifically, the feature extractor $ E $ is a convolutional neural network with eight convolutional layers to abstract latent feature representations from input samples. Furthermore, the on-off predictor $ P $ and the motion discriminator $ D $ are configured with three fully-connected layers to facilitate their own predictions.

\begin{figure*}
	\hfill
	\begin{minipage}[t]{0.32\linewidth}
		\centering
		\includegraphics[width=0.824\textwidth]{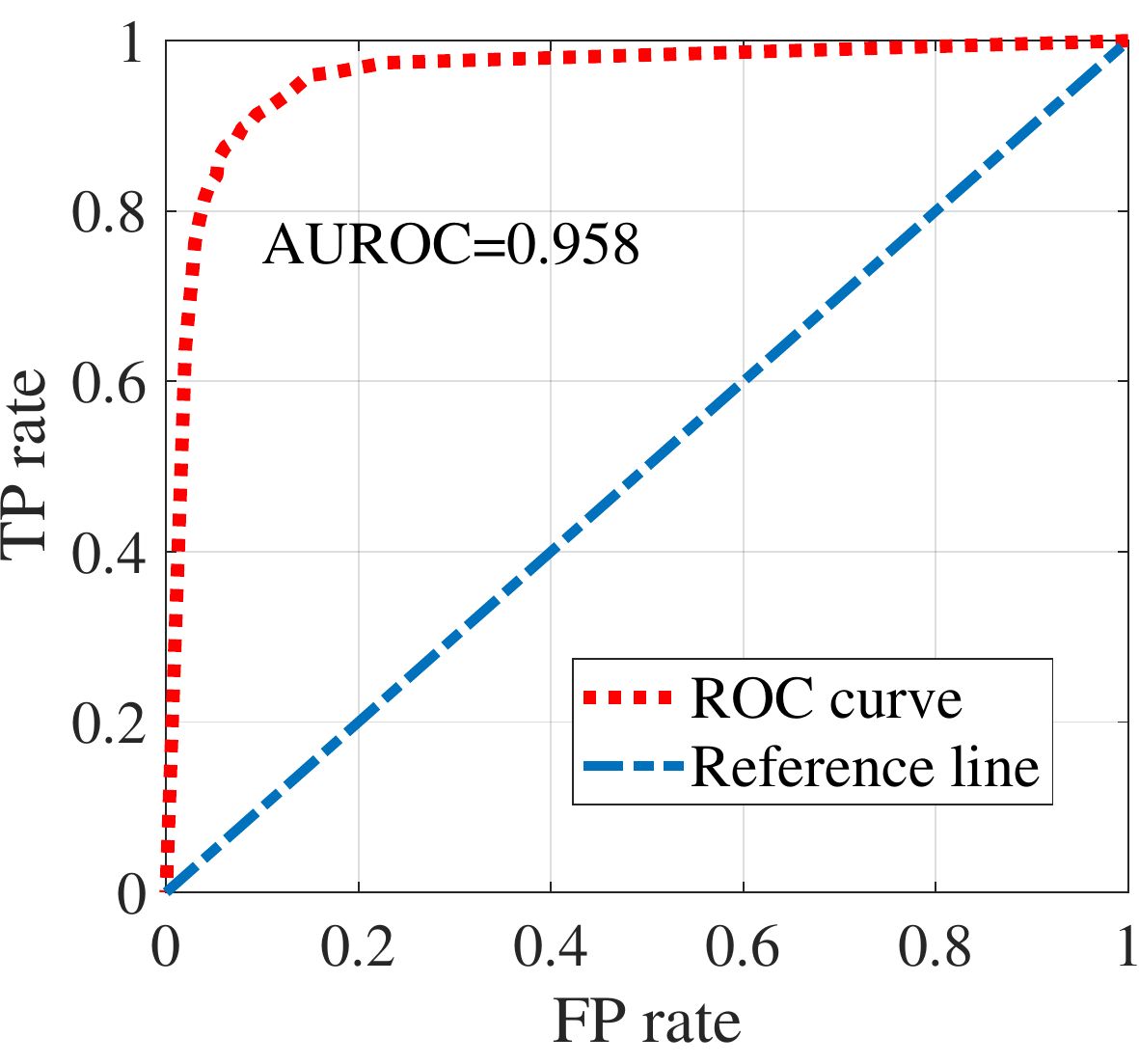}
		\caption{The ROC curve of the system. }\label{fig:roc}
	\end{minipage}
	\hfill
	\begin{minipage}[t]{0.32\linewidth}
		\centering
		\includegraphics[width=\textwidth]{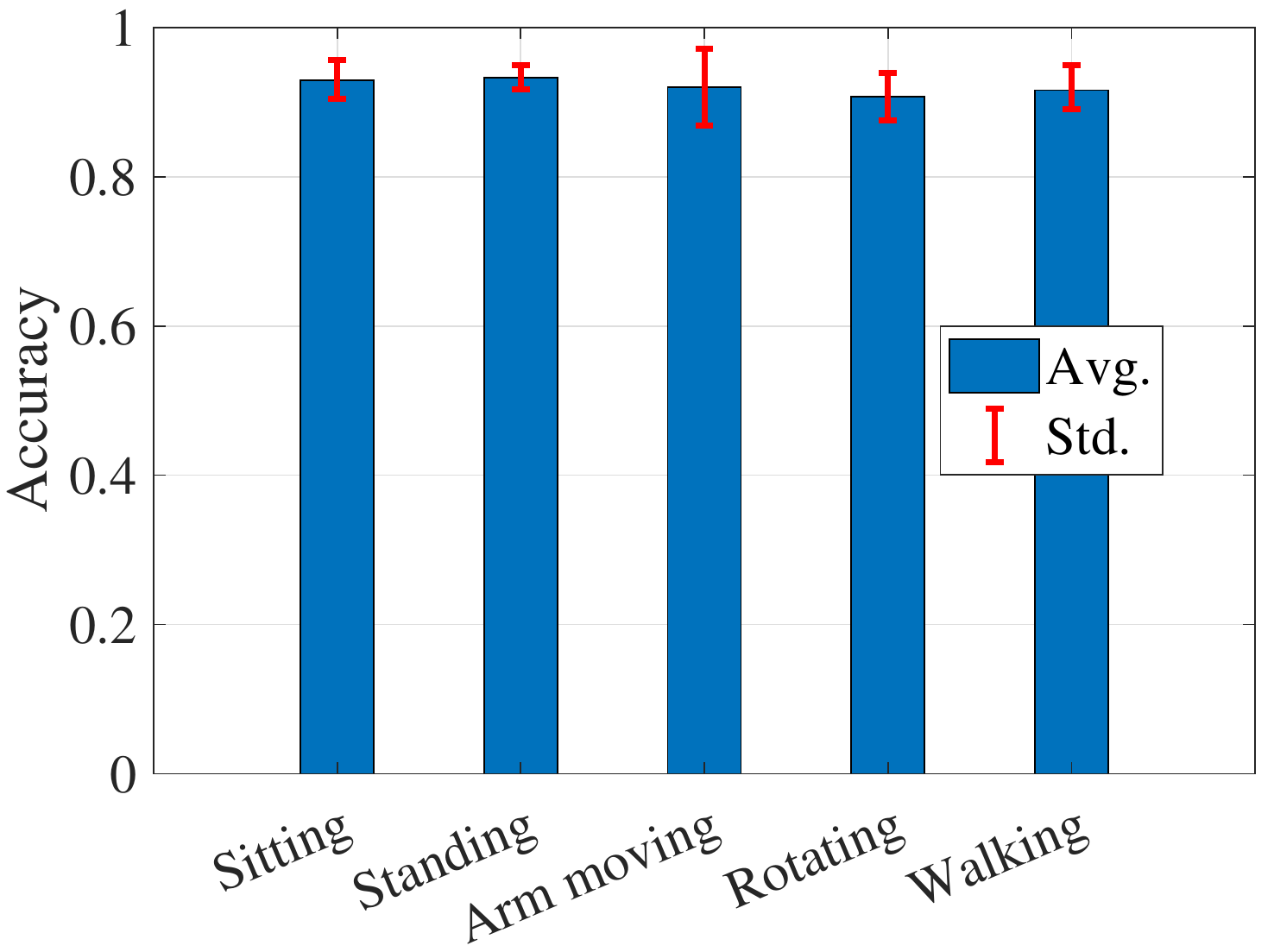}
		\caption{Accuracy for different body motions.}\label{fig:accuracy}
	\end{minipage}
	\hfill
	\begin{minipage}[t]{0.32\linewidth}
		\centering
		\includegraphics[width=\textwidth]{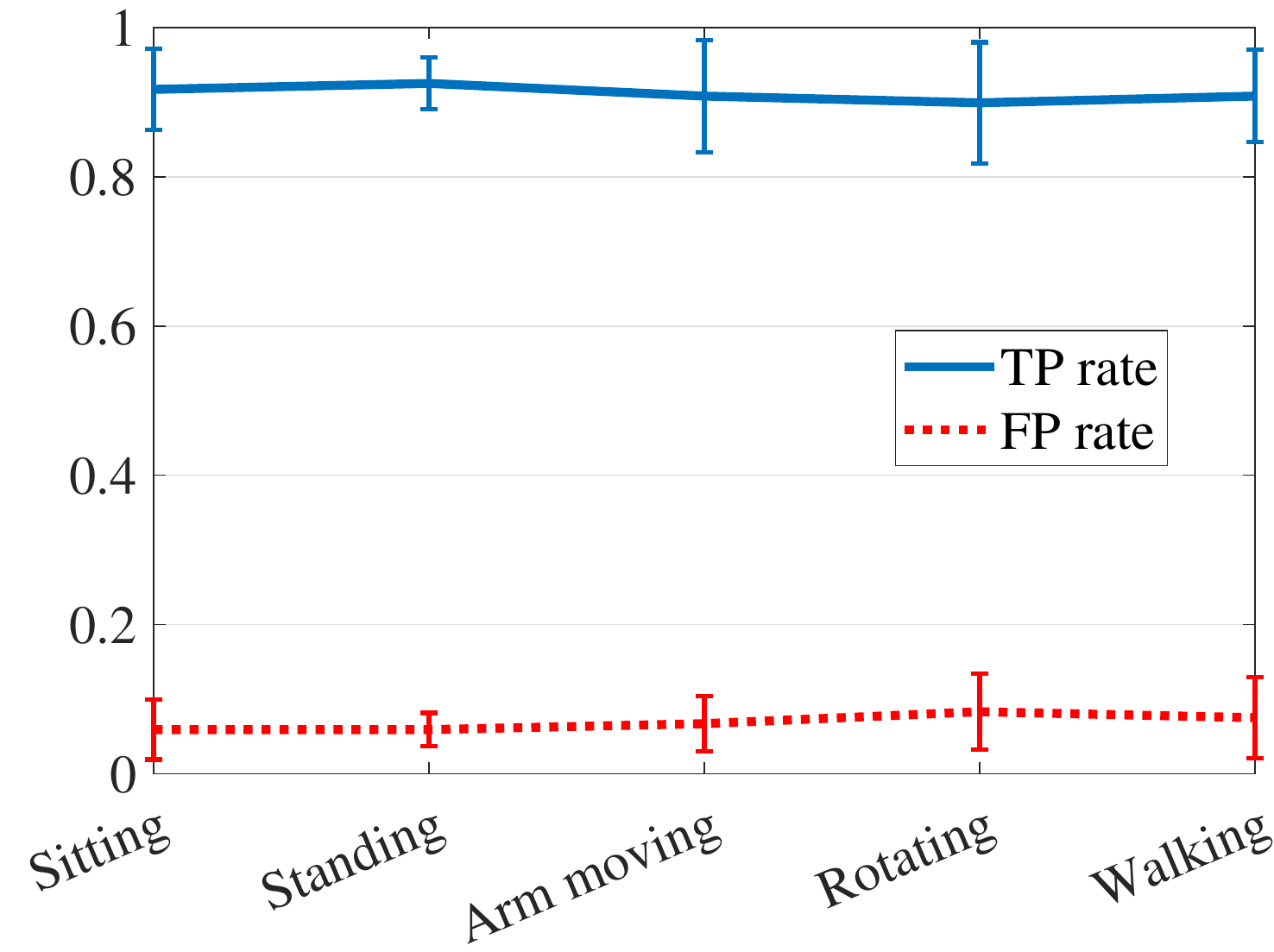}
		\caption{TP and FP rates for different body motions.}\label{fig:TP and FP rates}
	\end{minipage}
\end{figure*}

\textbf{Evaluation Metrics.} We use the following metrics to illustrate the performance of our system.
\begin{itemize}
	\item \textbf{Accuracy.} It is computed as the ratio of the number of RSS segments that are correctly recognized to the total number of on- and off-body RSS segments.
	\item \textbf{True positive (TP) rate.}  It is denoted as the ratio of the number of on-body RSS segments that are correctly predicted to the total number of on-body segments. 
	\item \textbf{False positive (FP) rate.} It is defined as the ratio of the number of off-body RSS segments that are mistakenly accepted to the total number of off-body segments.
\end{itemize}

\subsection{Performance Results} 
We first illustrate the overall performance of our authentication system on all testing data. As shown in Table~\ref{tab:overall_results}, our system is able to identify 90.4\% of on- and off-body devices on average. Specifically, it can correctly recognize on-body devices with a ratio of 89.0\% and successfully mitigate 91.8\% of attacks from off-body devices. In addition, we report the receiver operating characteristic (ROC) curve of our system, which depicts the tradeoff between FP and TP rates by varying their discrimination threshold in the interval $ \left[0,1 \right]  $. As depicted in Fig.~\ref{fig:roc}, the system's ROC curve first goes straight up and then becomes steady promptly as FP rate increases. Moreover, the area under the ROC curve (AUROC) reaches 0.958, which is close to 1, i.e., the AUROC of the ideal case. The above results indicate that our system achieves good authentication ability. 
\begin{table}[h]
	\centering
	\small
	\caption{Overall Performance Results}\label{tab:overall_results}
	\begin{tabular}{c|c|c}
		\hline
		\textbf{Accuracy} & \textbf{TP Rate} & \textbf{FP Rate} \\
		\hline
		90.4\% $ \pm $ 1.9\%  & 89.0\% $ \pm $ 2.4\%  &  8.2\% $ \pm $ 1.7\%\\
		\hline
	\end{tabular}
\end{table}

Then, we elaborate on the authentication performance for each frequently appearing motion. In general, each motion has a unique movement pattern of the human body, and thus exhibits different effects on BAN radio waves. As plotted in Fig.~\ref{fig:accuracy}, the proposed system achieves better performance for the static motions than for the dynamic ones. The same observations are also shown in Fig.~\ref{fig:TP and FP rates}. Therein, higher TP rates and lower FP rates are clearly present in the static states, because there are fewer disturbances caused by body movements in radio signals when the user sits or stands still with IoT devices, which makes it much easier for the system to recognize on- and off-body propagation patterns. Despite the above differences, the system still achieves average TP and FP rates of 90.8\% and 6.9\%, respectively, in the controlled user motion scenario.

\begin{figure*}
	\hfill
	\begin{minipage}[t]{0.295\linewidth}
		\centering
		\includegraphics[width=\textwidth]{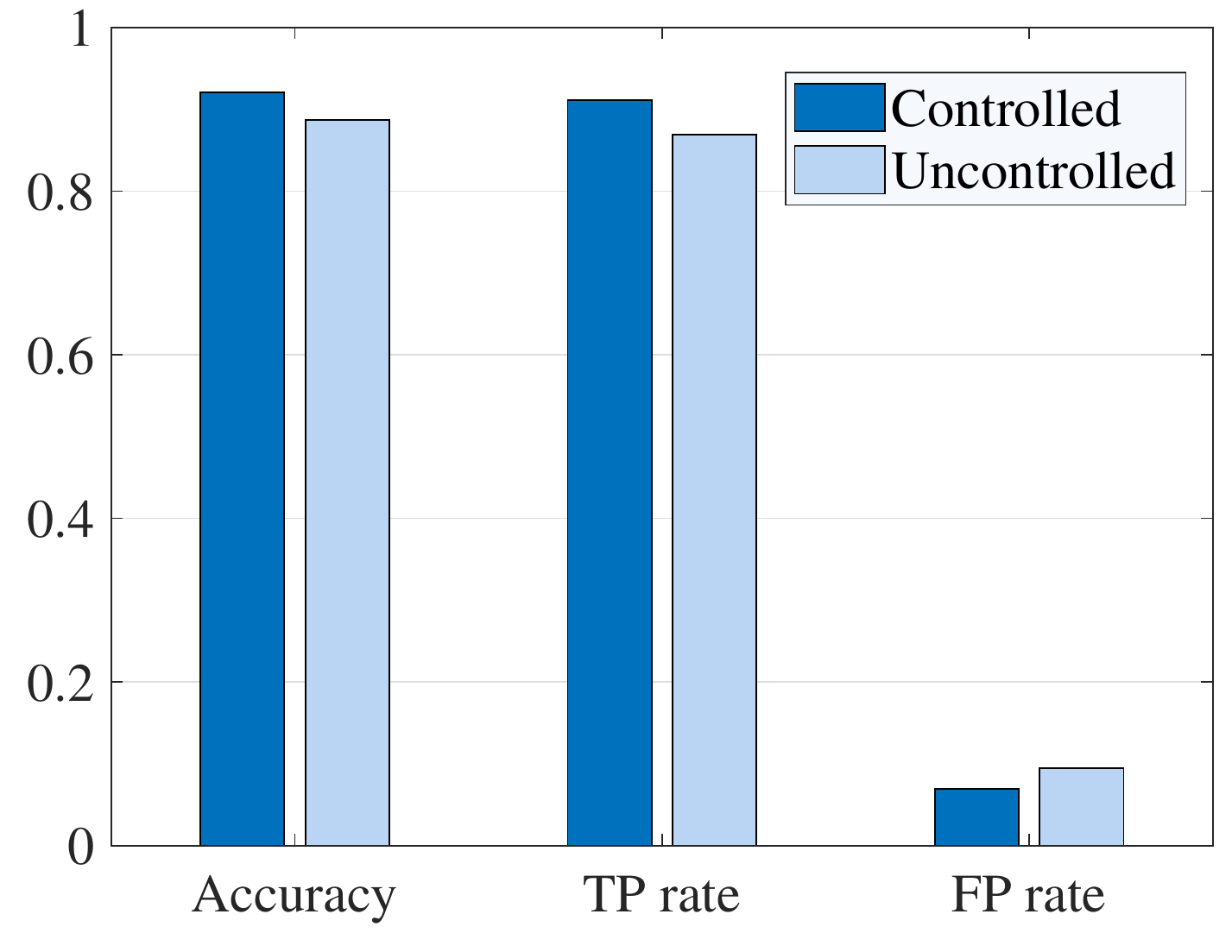}
		\caption{Performance comparison between the controlled and uncontrolled user motion scenarios.}\label{fig:performance overall}
	\end{minipage}
	\hfill
	\begin{minipage}[t]{0.31\linewidth}
		\centering
		\includegraphics[width=\textwidth]{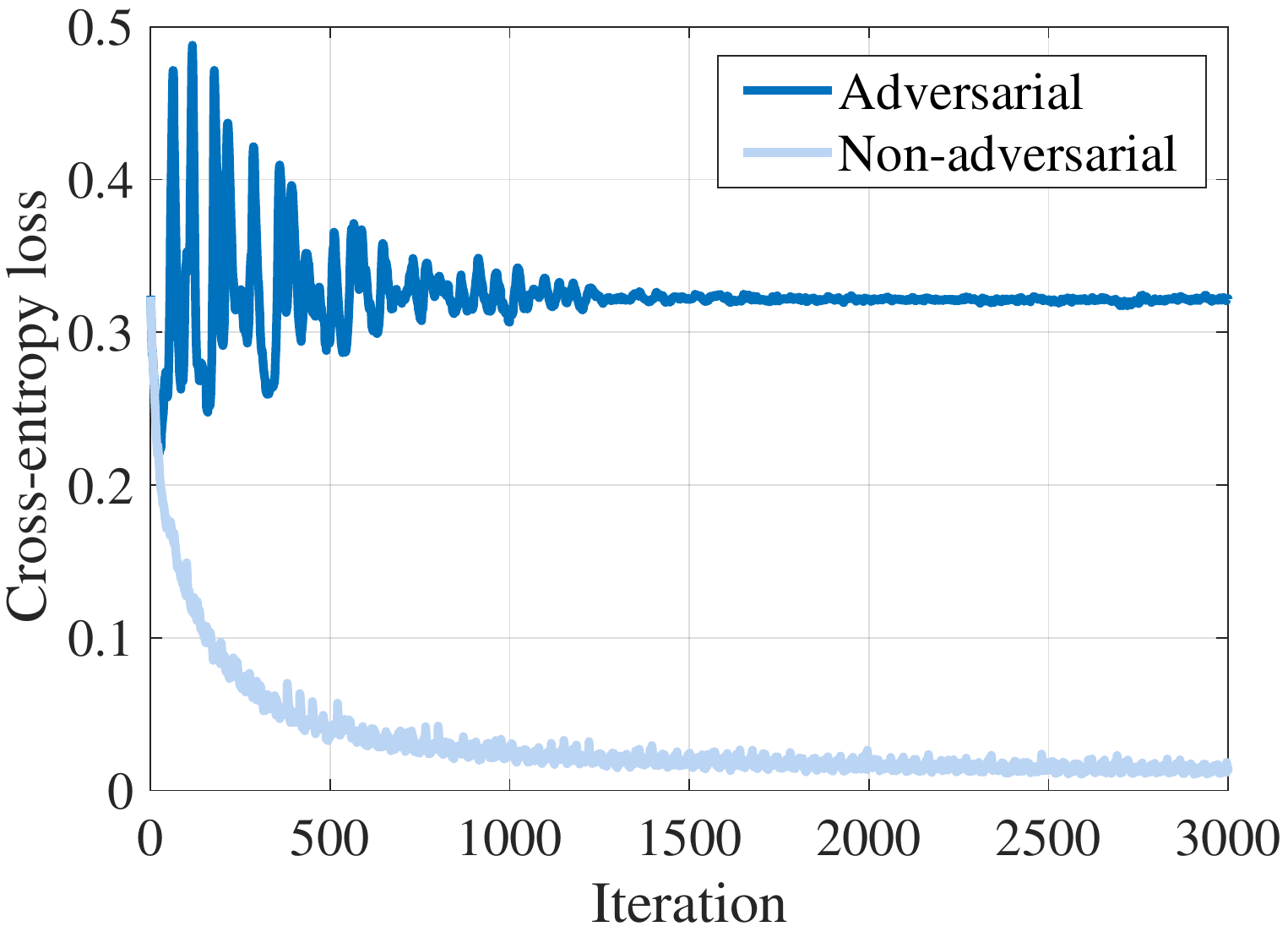}
		\caption{Training losses of discriminators for adversarial and non-adversarial models. The lower the value is, the more information pertaining to body motions a discriminator can learn.}\label{fig:losses for discriminator}
	\end{minipage}
	\hfill
	\begin{minipage}[t]{0.31\linewidth}
		\centering
		\includegraphics[width=\textwidth]{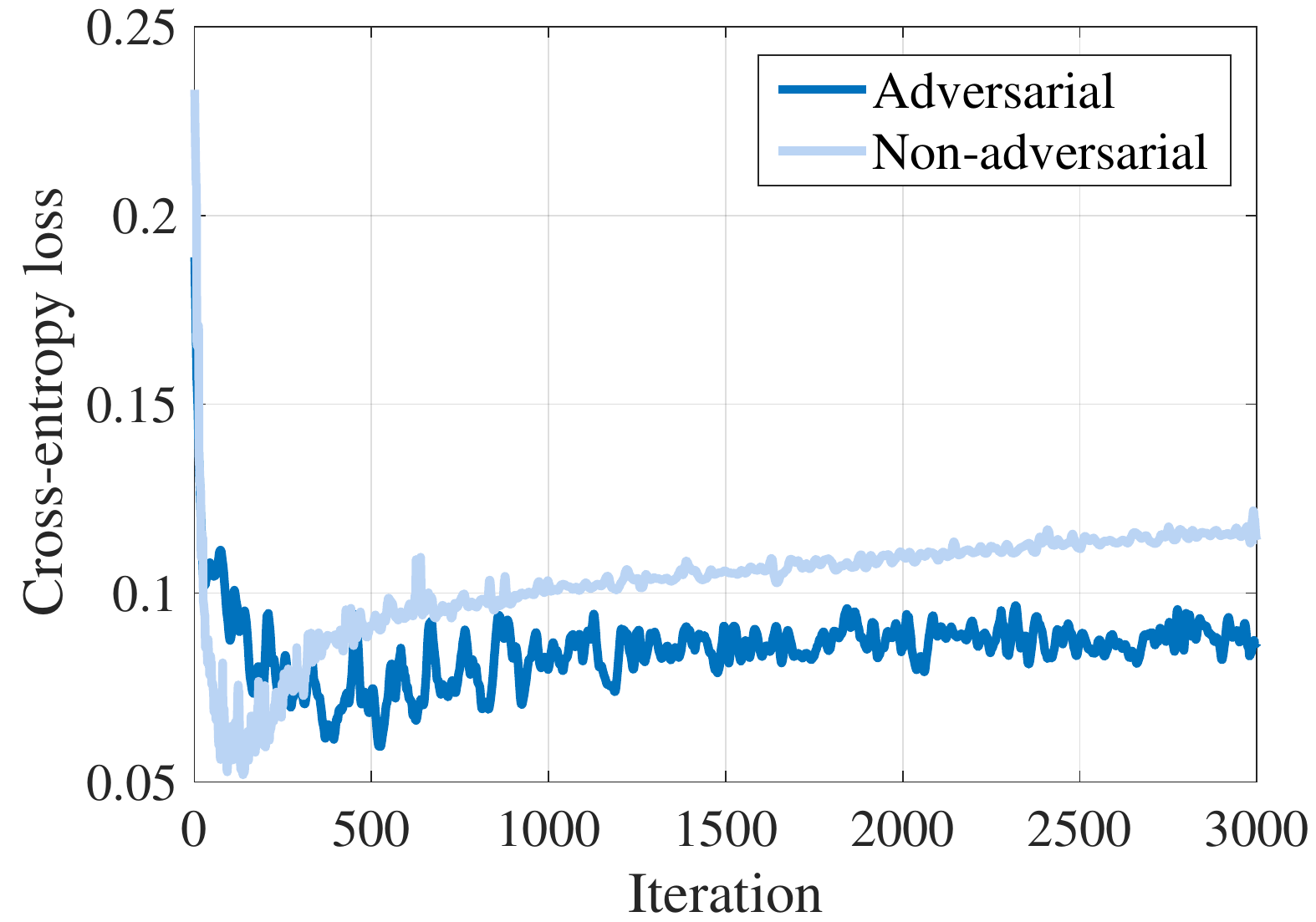}
		\caption{Testing losses of predictors for adversarial and non-adversarial models. The rise of loss curves is due to the over-fitting phenomenon.}\label{fig:losses for predictors}
	\end{minipage}
\end{figure*}

Next, we compare the system performance in the uncontrolled scenario with that in the controlled one. As illustrated in Fig.~\ref{fig:performance overall}, the system shows performance degradation in each metric in the uncontrolled scenario. The reason for the degradation is that more irregular and complicated body movements are present when the user behaves casually, which causes the extractor to extract more noisy features and thus hampers the prediction ability of the predictor. More specifically, the system has a TP rate reduction of 4.0\% and a FP rate increase of 2.6\% for uncontrolled motions. This is due to the fact that, compared with off-body signals, on-body signals, dominated by creeping waves, are more sensitive to user motion dynamics, which results in more on-body RSS segments to be mistakenly classified as off-body ones.

We further illustrate the benefits of adopting an adversarial discriminator in our multi-player model. Our discriminator aims at helping the extractor to discover transferable features and thus boosts the generalization ability of the predictor. To illustrate these merits, we set up a version of our model with a non-adversarial discriminator as a baseline. Note that in the baseline, the update of the extractor' parameters relies solely on the minimization of the predictor's loss.

Fig.~\ref{fig:losses for discriminator} plots the training losses of discriminators in our and baseline models. The loss of the non-adversarial discriminator declines quickly and then stabilizes at a very low level. However, ours first fluctuates dramatically and finally converges to a high value. This is due to the fact that at the beginning, the fluctuations of the adversarial loss are incurred by its minimax optimization, and they mitigate gradually as motion specific features irrelevant to the predictor fade out in the feature representation. The above observations reveal that the extractor in our model abstracts more transferable features than that in the baseline. Furthermore, comparing the performance of two predictors in Fig.~\ref{fig:losses for predictors}, we see that both loss curves decrease at first and then increase after certain numbers of iterations. However, the adversarial curve rises up at a lower speed than the non-adversarial one, which suggests that our adversarial discriminator works as a regularizer for alleviating over-fitting and enables the promotion of the predictor's generalization ability.

\section{Related Work}
Dedicated sensors, including accelerometers \cite{revadigar2017accelerometer}, bioimpedance sensors \cite{cornelius2014wearable}, motion sensors \cite{xu2016walkie} and capacitive touch sensors \cite{vu2012distinguishing}, have been used to differentiate on- and off-body devices. Additionally, various sensors in smartphones \cite{ren2013smartphone,das2014you} have been also exploited to identify devices or users. However, sensor-based approaches limit themselves to specified user motions or fitness related wearables.

Existing measurements \cite{di2011body,ryckaert2004channel} have shown that essential differences exist between on- and off-body radio propagations. Based on the above studies, radio propagation characteristics were examined to identify legitimate wearable devices \cite{wang2018securing}. In comparison to the prior work, our work develops a customized adversarial network to essentially extract underlying propagation patterns and obtains a better generalized authentication performance in various motion scenarios.

\section{Conclusion}
This paper presents a motion invariant authentication system to secure on-body IoT device pairing and data transmission by harnessing an adversarial multi-player network to effectively recognize underlying radio propagation patterns. Our system takes one step forward to embrace the advent of human-centric IoT by supporting various wearable devices under diverse user motions. Our theoretical analysis indicates that at equilibrium, our adversarial model is resilient to motion variances. We extensively evaluate the proposed system with various static and dynamic user motions in indoor and outdoor settings. The results shows that our system can recognize 89.0\% of legitimate devices while at the same time mitigating 91.8\% of impersonation attack attempts. 

\section*{Acknowledgement}
The work was supported in part by the NSFC under Grant 61871441, 91738202, 61729101, the RGC under Contract CERG 16203215, Young Elite Scientists Sponsorship Program by CAST under Grant 2018QNRC001, National Key R\&D Program of China under Grant 2017YFE0121500, the Key Laboratory of Dynamic Cognitive System of Electromagnetic Spectrum Space (Nanjing Univ. Aeronaut. Astronaut.), MIIT, China under Grant KF20181911.

\bibliographystyle{IEEEtran}
\bibliography{IEEEabrv,./Onbody_authentication}

\end{document}